# Monte Carlo Evaluation of FADE Approach to Anomalous Kinetics

## M. Marseguerra [1], A. Zoia [2]

*Department of Nuclear Engineering, Polytechnic of Milan, Via Ponzio 34/3, 20133 Milan, Italy*

## Abstract

In a wide range of transport phenomena in complex systems, the mean squared displacement of a particles plume has been often found to follow a nonlinear relationship of the kind $\langle x^2(t) \rangle \propto t^\alpha$, where $\alpha$ may be greater or smaller than *1*: these evidences have been described under the generic term of anomalous diffusion. In this paper we focus on subdiffusion, i.e. the case *0<α<1*, in presence of an external advective field. Widely adopted models to describe anomalous kinetics are Continuous Time Random Walk (CTRW) and its Fractional Advection-Dispersion Equation (FADE) asymptotic approximation, which accurately account for experimental results, e.g. in the transport of contaminant particles in porous or fractured media. FADE approximated equations, in particular, admit elegant analytical closed-form solutions for the particle concentration *P(x,t)*. To evaluate the relevance of the approximations which allow to derive the asymptotic FADE equations, we resort to Monte Carlo simulation (which may be regarded as an exact solution of the CTRW model): this comparison shows that the FADE equations represent a less and less accurate asymptotic description of the exact CTRW model as $\alpha$ becomes close to *1*. We propose higher order corrections which lead to modified integral-differential equations and derive new expressions for the moments of *P(x,t)*. These results are validated through comparison with those of Monte Carlo simulation, assumed as reference curves.

## 1. Introduction

A general approach to the analysis of transport phenomena is based on Continuous-Time Random Walk (CTRW) [15,19,22,23,35], in which the travel of a particle (a walker) in a medium is modelled as a series of jumps of random lengths, separated by random waiting times. The theory of CTRW with algebraically decaying probability distribution functions (pdf's) has been originally introduced in Physics in a series of seminal papers by Weiss, Scher, Montroll and co-workers [15,27,29,30,31] in the late 1960s to explain evidences of anomalous diffusion occurring in the drift-diffusion processes in amorphous semiconductors. The diffusion is called anomalous if the mean squared displacement (MSD) of the particles plume is not linearly proportional to time *t* as in the standard Fickian case, but to a power $\alpha$ of *t* larger (superdiffusion) or smaller (subdiffusion) than unity. More recently, anomalous diffusion has turned out to be quite ubiquitous in almost every field of science (see e.g. [22] and [23] for a detailed review) and the CTRW model has been applied with success to interpret the experimental results and to make predictions on the evolution of the examined systems. Such applications concern among the others e.g. the behaviour of chaotic Hamiltonian systems as related to the transport of charged particles in turbulent plasma [33,36], the evolution of financial markets [18], the dynamics of disordered systems [7] or the transport of contaminant particles in groundwater under the combined effect of rock fractures and porosity [3-6,8-10,16,20,21,25]. In this paper we will focus on the case of subdiffusion (*0<α<1*) on a 1D infinite support, in presence of an external (constant) advective field. In the subdiffusive case [3-6,8-10,16,19-21,22,25], an algebraically decaying distribution is assumed for the waiting times of the particles in the surrounding medium (instead of the traditional exponential one): this physically means that the particles will have a non-vanishing probability of extremely long sojourn times in the visited locations, due to so-called "trapping events" (see e.g. [36]). The macroscopic effect is that the variance of the particle plume grows less than linearly in time (subdiffusion). Within the CTRW approach to transport phenomena, advection can be modelled in either a Galilei invariant [22,25] or Galilei variant [3,4,20,22,25] scheme, each involving different hypotheses on the microscopic dynamics of the described particle plume. In general, the CTRW equations do not allow closed-form analytical solutions. However, if suitable first-order approximations are introduced in the Fourier and Laplace transforms expansions for both the jump lengths and waiting times distributions, Fractional Advection-Diffusion Equations (FADE) can be formally derived from the original exact CTRW model [22,23,32,33,34,36]. In this respect, FADE may be regarded as an asymptotic subset of CTRW. In order to evaluate the effects of the approximations introduced in the transformed space as to obtain the FADE equations, we resort to Monte Carlo simulation as a mean to accurately solve CTRW equation (as proposed e.g. in [3,10,21]): it turns out that relevant deviations of the analytical FADE results with respect to the Monte Carlo solutions are evident when $\alpha \to 1$. In particular, our analysis focuses on the moments of the distribution $P(x,t)$, which constitute accurate estimators of the solution accuracy [19,21]. These observations suggest that when $\alpha \to 1$ FADE is not an accurate asymptotic solution of CTRW and show the need of improving the first-order expansions which lead to the FADE equations in order to include higher-order terms: in this case, no analytical solutions are available and numerical Laplace inversion will be required to obtain the desired solutions of the CTRW equation and their moments. The proposed corrections to the transformed space expansions, validated by means of Monte Carlo simulation, show that the new set of derived equations can accurately approximate the behaviour of the exact CTRW solution even when $\alpha \to 1$.

---

[1] *Corresponding Author. Tel: +39 02 2399 6355. Fax: +39 02 2399 6309*
*Email address: marzio.marseguerra@polimi.it (Marzio Marseguerra)*

[2] *Email address: andrea.zoia@polimi.it (Andrea Zoia)*

The paper is organized as follows: in Section 2 we summarize the basic concepts of the CTRW approach and the derivation of the Fractional Diffusion Equation (FDE). In Section 3 we present the Galilei invariant and variant schemes within the CTRW approach and we derive the corresponding FADE: analytical results of the asymptotic FADE will be compared to Monte Carlo simulation results. We will show that discrepancies arise when $\alpha \to 1$. In Section 4 we propose higher order corrections for both advection schemes and compare the numerical results to the Monte Carlo simulation findings. Conclusions are finally drawn in Section 5. Appendices A and B are devoted to the required calculations.

## 2. CTRW and FDE approaches to subdiffusion

Let $X(t)$ be a stochastic process describing the motion of a tracer particle (a walker) performing random jumps separated by random waiting times. The associated probability density function (pdf) $P(x,t)$ represents the probability of the walker being in $X=x$ at time $t$ and it is also called the propagator of the process. The CTRW approach to the description of this stochastic process is based on the probability balance expressed by the Chapman-Kolmogorov integral equation (the so-called Master Equation), which entails the pdf $P(x,t)$ [22,35,36]. It can be shown [12,18,22,23,36] that, if $\lambda(x)$ and $w(t)$ are the pdf's of the single jump lengths and waiting times[1], respectively, the Laplace- and Fourier- transformed expression of $P(x,t)$ satisfies the simple algebraic relation

$$P(k,u) = \frac{1-w(u)}{u} \frac{1}{1-w(u)\lambda(k)} \qquad (1)$$

in the case of a Cauchy problem with initial conditions $\delta(x)\delta(t)$. For convenience, adopting a well-established convention (see e.g. [19,22,23]), we denote the Fourier or Laplace transform of a pdf by its argument, namely $\Im\{f(x);k\} = f(k)$ and $\mathcal{L}\{f(t);u\} = f(u)$. $P(k,u)$ is called the propagator of the underlying stochastic process: equation (1), in the doubly transformed space, represents a probability (normalized mass) balance for the number of particles and in this respect may be regarded as an exact transport formulation. According to the choice of the waiting times and of the jump lengths distributions, the CTRW approach may give rise to (normal) diffusive, subdiffusive and superdiffusive behaviour of the walker. For the case of subdiffusion, i.e. *0<α<1*, the most widely adopted choice for $\lambda(x)$ is a Gaussian pdf with finite variance, while $w(t)$ is assumed to be any algebraically decaying pdf of the kind $w(t) \sim t^{-1-\alpha}$ when $t \to +\infty$. The specific functional form of $w(t)$, whose first moment is infinite, introduces "memory effects" in the trajectory $x = x(t)$ of the walker, giving rise to long range correlations which make its path semi-Markovian[2]. This fact in turn prevents the application of the Central Limit Theorem and results in anomalous diffusion, i.e. $\langle x^2(t) \rangle \propto t^\alpha$ [11,18,22,36]. In general, as mentioned before, no analytical solution is known for (1), as the required Laplace and Fourier inverse transforms are usually not trivial[3]. However, if we consider the asymptotic behavior of the solution sufficiently far from the origin, i.e. the so-called "diffusion limit" approximation $\{|x| \to +\infty, t \to +\infty\} \leftrightarrow \{|k| \to 0, u \to 0\}$ and expand the Laplace and Fourier transforms $\lambda(k)$ and $w(u)$ up to the first non-constant term, the required inverse transforms may be explicitly performed. Therefore, from (1) by means of the definition of the differential operators in the transformed spaces we can formally derive a Fractional Diffusion Equation (FDE) whose analytical solution is known and may be expressed in terms of the Fox $H$ function, as first shown by Schneider and Wyss [17,22,26,32].

---

[1] Here, for sake of simplicity, we adopt the hypothesis that the jump length and waiting time probabilities are independent. In general, however, they may well be correlated.
[2] See below and Appendix B for details on the semi-Markovian nature of the walker when *α<1*.
[3] It must be mentioned here that it is possible to resort to a rephrasing of the original probability balance Master Equation in terms of an ordinary differential equation which can be then analytically solved with respect to space to get P(x,u). However, to obtain the full solution of the CTRW equation P(x,u) needs to be numerically Laplace inverted with respect to time [8-10].

In the following, for sake of simplicity we will assume a pdf for the waiting times of the kind

$$w(t) = A \left.\frac{t}{\tau^2}\right|_{t \leq \tau} + A \left.\frac{\tau^\alpha}{t^{1+\alpha}}\right|_{t > \tau} \tag{2}$$

where $A$ is a normalization factor, which reads $A = \dfrac{2\alpha}{2+\alpha}$, and $\tau$ is a characteristic time constant. This pdf has an asymptotic behaviour $w(t) \approx \dfrac{1}{t^{1+\alpha}}$ when $t \to +\infty$, so that we expect $w(t)$ to converge to a one-sided Lévy stable distribution of index $\alpha$ [11]. Indeed, if the Laplace transform of (2) is expanded in the long time limit (i.e. $u\tau \ll 1$) and truncated to the first non-constant term in $u$ (details are left to Appendix A), we get

$$w(u) \approx 1 - c_\alpha (u\tau)^\alpha, \tag{3}$$

where $c_\alpha = \dfrac{2\Gamma(1-\alpha)}{2+\alpha}$.

Equation (3) represent the expansion for small $u$ of a Laplace transformed one-sided Lévy stable distribution $L(u) = e^{-c_\alpha (u\tau)^\alpha}$ [11,22].

On the other hand, if we further assume that the jump lengths distribution $\lambda(x)$ is a Gaussian pdf with mean $\mu = 0$ and variance $\Sigma^2 = 2\sigma^2$, then its Fourier transform $\lambda(k)$ in the diffusion limit expansion $\kappa\sigma \ll 1$, i.e. far from the spatial origin, becomes

$$\lambda(k) \approx 1 - \sigma^2 \kappa^2 \ . \tag{4}$$

If expansions (3) and (4) are substituted in (1), after some algebra and recalling the properties of the Riemann-Liouville operator[1] ${}_0\partial_t^{1-\alpha}$, the propagator (1) may be formally reverted to the $\{x,t\}$ space, to finally obtain

$$\frac{\partial}{\partial t} P(x,t) = D_\alpha \, {}_0\partial_t^{1-\alpha} \frac{\partial^2}{\partial x^2} P(x,t) \tag{5}$$

where $D_\alpha = \dfrac{\sigma^2}{c_\alpha \tau^\alpha}$ may be thought as a generalized diffusion coefficient. Indeed, equation (5), which assumes the name of Fractional Diffusion Equation, is a generalization of the classical Fickian Diffusion Equation (DE) and describes the asymptotic behaviour of a plume of subdiffusive particles in the absence of advective contributions to the motion[2]. As mentioned before, the presence of the Riemann-Liouville operator in (5) accounts for slowly decaying time correlations and memory effects[3] which slow down the particles motion: as such, it is responsible for the semi-Markovian nature of the particles kinetics [18,34,36], in contrast with the Markovian nature of the normal diffusive particles as described by the standard diffusion equation. We shall now discuss the distinct merits and drawbacks of the fractional-in-time formulation (5) with respect to the original CTRW (1). As for its drawbacks, FDE arises as an asymptotic approximation of a more general and exact transport model, formulated in terms of the Chapman-Kolmogorov Master Equation, as said before (see e.g. [2] for a detailed discussion). Moreover, while anomalous diffusion is often experimentally found to be a transient phase, which – after a suitable time interval – generally relaxes towards Fickian diffusion, FDE can not to take into account this transition in a straightforward manner, since the anomalous behaviour is assumed to hold even at infinite time [9,10,22,23,32]. On the other hand, a prominent advantage of FDE with respect to the CTRW approach is that the fractional derivative formulation may easily include external fields in a simple manner and it is naturally suitable to solve boundary value problems [22-24]. In this respect, FDE has been recently shown to act as a unifying framework for the quantitative description of different physical phenomena where anomalous diffusion plays a significant role [22,23]. Moreover, a plethora of standard mathematical techniques derived from partial differential equations literature are readily available to obtain analytical solutions for FDE. These considerations on FDE model are equally valid also for the case of FADE, which we will discuss in the following Sections.

---

[1] By definition, ${}_0\partial_t^{1-\alpha}\{f(x,t)\} \equiv \dfrac{1}{\Gamma(\alpha)} \dfrac{\partial}{\partial t} \int_0^t \dfrac{f(x,t')}{(t-t')^{1-\alpha}} dt'$ for any sufficiently well-behaved function $f$ and $0<\alpha<1$. It follows that in the Laplace transformed space $\mathscr{L}\{{}_0\partial_t^{-\alpha}(f(x,t))\} = u^{-\alpha} f(x,t)$, $\alpha \geq 0$ [17,22,26,28].

[2] A more general fractional differential equation was derived from CTRW in [1] and [36].

[3] See Appendix B for further details.

We end this Section by remarking that the accuracy of the asymptotic solutions of (5) with respect to the exact CTRW formulation has been discussed elsewhere [3,21]. As this point will be central in the development of next Sections, here we mention that when $\alpha \to 1$ the moments of distribution $P(x,t)$ show neat discrepancies with respect to those of the solution of (1) computed via Monte Carlo simulation (by sampling waiting times from the exact pdf (2), without resorting to any approximation), thus suggesting that the FDE (5) accurately represents the asymptotic behaviour of (1) only when $\alpha$ is small, say $\alpha<0.5$: when $\alpha \to 1$ an higher order term in the expansion of $w(u)$ comes into play and its growing importance may be evaluated by means of Monte Carlo simulation. This is in agreement with the findings of [19] for the case of subdiffusive transport in presence of a forward bias, for which analogous corrections to $w(u)$ have been shown to be strictly required.

## 3. Modelling an advection field within CTRW approach

Once the general form of the propagator (1) has been set, there basically exist two schemes which allow to take into account an external velocity field in presence of subdiffusion: this distinction has been first introduced in [25]. Here, for sake of simplicity, we will consider a constant homogeneous advective field. The first scheme is introduced by requiring that the solution $P(x,t)$ of (1) be invariant under a transformation of coordinates of the kind $x \to x - vt$: this idea stems directly from the standard Fickian Advection-Dispersion Equation (ADE). This scheme, which is called "Galilei invariant" because of the mentioned invariance property, assumes that in the moving frame (at velocity $v$) each particle is ruled by the usual pdf's $\lambda(x)$ and $w(t)$. Galilei invariant subdiffusion may be found e.g. for tracer particles immersed in polymer solutions, where the flowing substance itself is the cause of the slowing down of the motion [22,25]. If we indicate $\varsigma(x,t)$ and $\chi(x,t)$ the jump probability in the moving frame and in the laboratory frame, respectively, then the assumption of Galilei invariance implies that

$$\chi(x,t) = \varsigma(x - vt, t), \text{ or equivalently } \chi(k,u) = \varsigma(k, u + ivk) \quad [22,25]. \tag{6}$$

If we assume expansions (3) and (4) to hold true, propagator (1) readily becomes

$$P(k,u) = \frac{1}{u + ivk + D_\alpha u^{1-\alpha} k^2} \tag{7}$$

Again, recalling the properties of ${}_0\partial_t^{1-\alpha}$ in the transformed space and with the help of some algebra, propagator (7) may be formally inverted to the $\{x,t\}$ space to give

$$\frac{\partial}{\partial t} P(x,t) + v \frac{\partial}{\partial x} P(x,t) = D_\alpha \, {}_0\partial_t^{1-\alpha} \frac{\partial^2}{\partial x^2} P(x,t) \tag{8}$$

which is called the (Galilei invariant) Fractional Advection Dispersion Equation (FADE) and may be considered as a generalization of the well-known ADE. Its analytical solution may be obtained in terms the solution of (5) by observing that by definition $P(x,t;v) = P(x - vt, t; v = 0)$. Moreover, recalling that by definition

$$\left\langle x^n(t) \right\rangle = \mathcal{L}^{-1} \left\{ \lim_{k \to 0} (i^n) \frac{\partial^n}{\partial k^n} P(k,u) \right\}, \tag{9}$$

from equation (7) it is possible to deduce all the moments of $P(x,t)$: in particular, under approximations (3) and (4), the Laplace inverse transform in (9) can be analytically computed, to obtain

$$\left\langle x^1(t) \right\rangle = vt, \tag{10}$$

$$\langle x^2(t)\rangle = \frac{2D_\alpha}{\Gamma(1+\alpha)} t^\alpha + (vt)^2, \qquad (11)$$

so that $\langle (x(t)-\langle x(t)\rangle)^2\rangle = \dfrac{2D_\alpha}{\Gamma(1+\alpha)} t^\alpha$. (12)

The moments of distribution *P(x,t)* are sensitive estimators of the accuracy of *P(x,t)* in (8) as an approximation of the exact solution of (1). Since, as required by the Galilei invariance property, the variance (12) has exactly the same expression of the one of the FDE without advection [22,25], the accuracy of the FADE (8) is traced back to the one of the FDE and we expect expressions (11) and (12) to be accurate when $\alpha<0.5$ and to show relevant discrepancies as $\alpha \to 1$. The first moment (10) is not influenced by the order of expansion (3). In the following Figures for two different values of $\alpha$ we compare the analytical approximate variance (12) with the one computed by resorting to Monte Carlo simulation as described in [21]. The simulation parameters are the following: $10^5$ particles have been followed up to a time $t_{max}= 10^5$, with $\sigma$=0.5, $\tau$=1 and *v*=1 when $\alpha$=0.5 (Figure 1). Then we set $\sigma$=0.5, $\tau$=1, *v*=1 and $t_{max}$=$10^4$ when $\alpha$=0.8 (Figure 2). We remark that in both cases the explored time scales guarantee that $\tau$ is "small", thus ensuring that the diffusion limit is a justified assumption. As expected, for large $\alpha$ FADE (8) is not an accurate approximation of (1) (Figure 2).

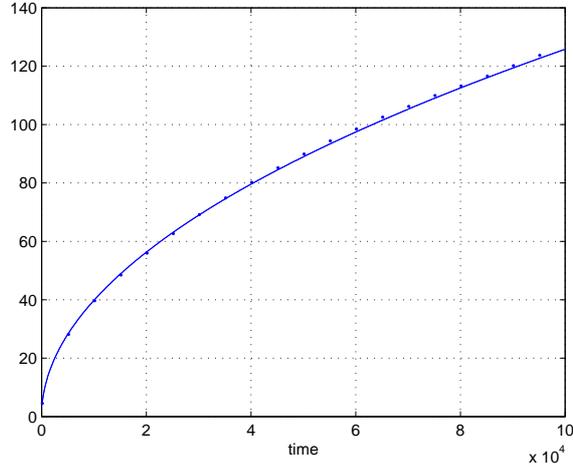

*Figure 1. Comparison between analytical variance (12) (solid line) and Monte Carlo variance (dots) when $\alpha$=0.5.*

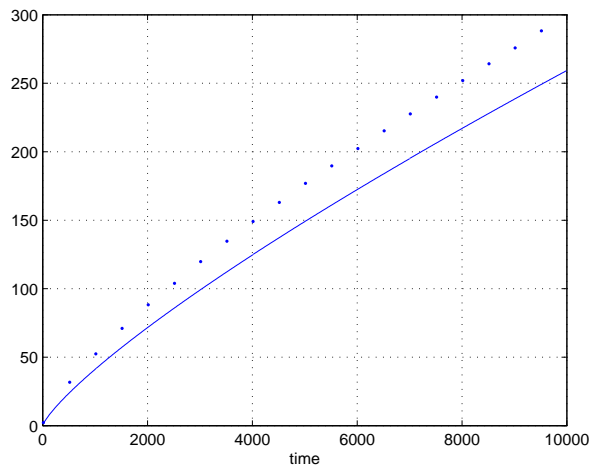

*Figure 2. Comparison between analytical variance (12) (solid line) and Monte Carlo variance (dots) when $\alpha$=0.8.*

We proceed now further to examine the other possible scheme to model an advection field within the CTRW approach, namely the Galilei variant scheme. Starting again from (1), it is assumed that each jump has a spatial (constant) bias in the forward direction, $\mu$. This assumption does not result in a Galilei invariance property for *P(x,t)*, hence the name of the scheme [25]. Experimental evidences of this kind of behaviour may be e.g. found in the context of flow through porous or fractured media (and in general in disordered systems) [3,4,19,20,22,25]. This scheme is modelled by adding to the jump lengths pdf $\lambda(x)$ a finite mean $\mu$. Thus, the expansion in small $k$ of the Fourier transform (4) becomes

$$\lambda(k) \approx 1 + i\mu k - \sigma^2 k^2. \tag{13}$$

We remark that $\mu$ has the dimensions of a space. Assuming that expansion (3) still holds true, one readily deduces from (1) the propagator

$$P(k,u) = \frac{1}{u + D_\alpha k^2 u^{1-\alpha} + iv_\alpha k u^{1-\alpha}}, \tag{14}$$

where $v_\alpha = \frac{\mu}{c_\alpha \tau^\alpha}$ is a generalized advection coefficient (or velocity). Again, by taking into account the properties of the Riemann-Liouville operator, propagator (14) may be formally inverted in the $\{x,t\}$ space, to finally obtain

$$\frac{\partial}{\partial t} P(x,t) = {}_0\partial_t^{1-\alpha} \left( D_\alpha \frac{\partial^2}{\partial x^2} P(x,t) - v_\alpha \frac{\partial}{\partial x} P(x,t) \right). \tag{15}$$

Equation (15) is called (Galilei variant) FADE or Fractional Fokker-Plank-Kolmogorov Equation (FFPK) [18,33,36]: its structure is neatly different from that of the Galilei invariant FADE, because of the presence of the fractional differential operator acting also on the advective contribution. Now, by making use of definition (9), we can derive all the moments of the distribution $P(x,t)$ solution of (15): their comparison with those computed by means of Monte Carlo simulation will provide an indication on the accuracy of the asymptotic $P(x,t)$ with respect to the exact solution of CTRW (1). After some algebra, we obtain:

$$\langle x^1(t) \rangle = v_\alpha \frac{t^\alpha}{\Gamma(1+\alpha)} \tag{16}$$

$$\langle x^2(t) \rangle = \frac{2 D_\alpha}{\Gamma(1+\alpha)} t^\alpha + \frac{2(v_\alpha)^2}{\Gamma(1+2\alpha)} t^{2\alpha}. \tag{17}$$

In the following figures for two different values of $\alpha$ we compare the analytical approximate variance $\langle (x(t) - \langle x(t) \rangle)^2 \rangle$ with the one computed by resorting to Monte Carlo simulation as described in [21]. The simulation parameters are the following: $10^5$ particles have been followed up to a time $t_{max} = 10^5$, with $\sigma=0.5$, $\tau=1$ and $\mu=0.2$ when $\alpha=0.5$ (Figure 3). Then we set $\sigma=0.5$, $\tau=1$, $\mu=0.2$ and $t_{max}=10^4$ when $\alpha=0.8$ (Figure 4). As expected, for large $\alpha$ FADE (15) is not an accurate approximation of (1) (Figure 4).

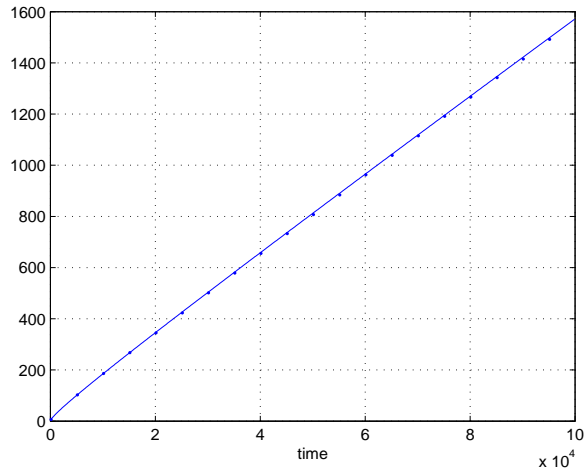

*Figure 3 Comparison between analytical (solid line, from (16) and (17)) and Monte Carlo variance (dots) when $\alpha=0.5$.*

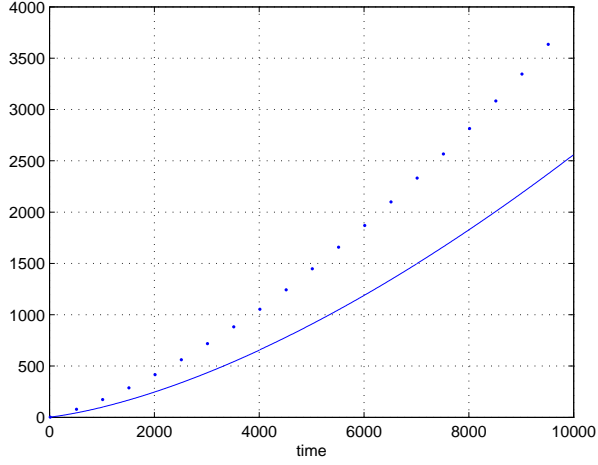

*Figure 4. Comparison between analytical (solid line, from (16) and (17)) and Monte Carlo variance (dots) when $\alpha=0.8$.*

**4. Higher order expansion of Laplace transform (3)**

In reason of the considerations presented in the previous paragraphs and elsewhere [19,21], we can attribute the discrepancies shown in Figures 2 and 4 to the truncation to the first non-constant term in (3), which in turn have led to the FADE formulation. Indeed, it can be shown that the Laplace transform expansion of pdf (2) in the diffusion limit $u\tau \ll 1$ and expanding up to the second order reads

$$w(u) \approx 1 - c_\alpha (u\tau)^\alpha + c_1 (u\tau) + o(u^2), \qquad (18)$$

where $c_1 = \left[ \dfrac{2\alpha}{(2+\alpha)(1-\alpha)} - \dfrac{2}{3}\dfrac{\alpha}{2+\alpha} \right]$. We refer the reader to Appendix A for the calculations details. When $\alpha$ is small, in the limit $u\tau \ll 1$ the linear term in $u$ in (18) is expected to play no significant role, thus justifying the accuracy of the FADE equations as approximations of the solution of (1). On the other hand, when $\alpha \to 1$ the two terms of expansion (18) become comparable and the effects of the linear contribution will no more be negligible. This gives an intuitive picture of the observed discrepancies between the FADE approach and the Monte Carlo solutions of CTRW, which by construction are not truncated to a first order expansion. Nonetheless, we must remark that the contribution of the linear term in (18) is expected to be completely negligible also for $\alpha$ close to 1 provided that the time constant $\tau$ is "sufficiently" small, by the meaning itself of expansion (18) as compared to (3). However, in realistic applications of anomalous diffusion, the value of $\tau$ is usually imposed by experimental evidences, so that a priori it is interesting to systematically explore the behaviour of the CTRW and FADE solutions in the intermediate range of values of $\tau$ for which the diffusion limit expansion $u\tau \ll 1$ holds true but the truncation of (18) to the first non-constant term might be inappropriate, depending on the value of the parameter $\alpha$. In order to assess the validity of the proposed second order corrections for (3), we replace expansion (3) with (18) in both propagators (7) and (14), to obtain:

$$P(k,u) = \dfrac{1}{u + ivk + \dfrac{\sigma^2 k^2}{c_\alpha \tau^\alpha u^{\alpha-1} - c_1 \tau}} \qquad (19)$$

for the Galilei invariant case and

$$P(k,u) = \dfrac{1}{u + \dfrac{\sigma^2 k^2}{c_\alpha \tau^\alpha u^{\alpha-1} - c_1 \tau} + \dfrac{i\mu k}{c_\alpha \tau^\alpha u^{\alpha-1} - c_1 \tau}} \qquad (20)$$

for the Galilei variant case, respectively. Consequently, we can then derive the integral-differential equations corresponding to the propagators (19) and (20) in the $\{x,t\}$ space. If we define $\Psi(u) = \dfrac{u\,w(u)}{1 - w(u)}$, where $w(u)$ is given by expansion (18), it can be shown that (19) is formally equivalent to the transport equation

$$\frac{\partial}{\partial t}P(x,t)+v\frac{\partial}{\partial x}P(x,t)=\int_0^t \Psi(t-t')\left(\sigma^2 \frac{\partial^2}{\partial x^2}P(x,t')\right)dt' \tag{19a}$$

and that (20) is formally equivalent to the transport equation

$$\frac{\partial}{\partial t}P(x,t)=\int_0^t \Psi(t-t')\left(\sigma^2 \frac{\partial^2}{\partial x^2}P(x,t')-\mu\frac{\partial}{\partial x}P(x,t')\right)dt' \tag{20a}$$

Detailed derivation of equations (19a) and (20a), together with some comments on the physical meaning of $\Psi(u)$, are left to Appendix B. Now, by resorting to definition (9) it is possible to derive the corresponding modified expressions of the moments of the two distributions: in particular, we obtain:

$$\left\langle \left(x(t)-\langle x(t)\rangle\right)^2 \right\rangle = \mathcal{L}^{-1}\left\{\frac{2\sigma^2}{u^2}\frac{1}{c_\alpha \tau^\alpha u^{\alpha-1}-c_1\tau}\right\} \tag{21}$$

for the variance of (19) and

$$\langle x(t)\rangle = \mathcal{L}^{-1}\left\{\frac{1}{u^2}\frac{\mu}{c_\alpha \tau^\alpha u^{\alpha-1}-c_1\tau}\right\} \tag{22}$$

$$\langle x^2(t)\rangle = \mathcal{L}^{-1}\left\{\frac{2}{u}\frac{\mu^2}{\left(c_\alpha \tau^\alpha u^\alpha - c_1\tau u\right)^2}\right\} + \mathcal{L}^{-1}\left\{\frac{2\sigma^2}{u^2}\frac{1}{c_\alpha \tau^\alpha u^{\alpha-1}-c_1\tau}\right\} \tag{23}$$

for the first and second moments of (20). The inverse Laplace transform appearing in (21-23) may be computed with a numerical algorithm [13-14] and the obtained curves are compared with those obtained from the Monte Carlo simulations described above: the next Figures 5 and 6 show the improved accuracy of the second order asymptotic equations (21-23) with respect to expressions (12,16-17). The accuracy of the proposed higher order corrections of expansion (3) are fully validated by the comparison with the Monte Carlo results, which may be regarded as a reference solution for the CTRW, no approximation having been introduced in the stochastic modelling, but for finite statistics effects.

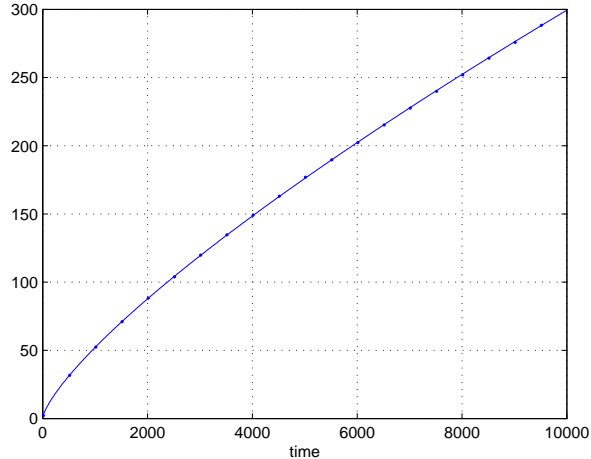

*Figure 5. Comparison between numerically inverted variance (21) and Monte Carlo variance (dots) when $\alpha=0.8$.*

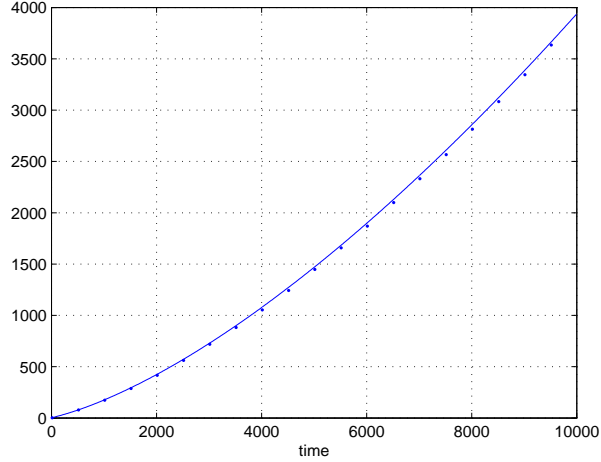

*Figure 6. Comparison between numerically inverted variance (from (22) and (23)) and Monte Carlo variance (dots) when $\alpha=0.8$.*

## 5. Conclusions

In a wide range of complex systems, the relaxation following an initial pulse is experimentally found to follow a nonlinear relationship of the kind $\langle x^2(t) \rangle \propto t^\alpha$, where $\alpha$ may be greater or smaller than *1*: these evidences have been described under the generic term of anomalous kinetics. In particular, in this paper we have focused on the case of subdiffusion, i.e. *0<α<1*. In order to properly characterize the time evolution of such systems, several approaches have been proposed in literature, among which the CTRW model has been shown to be particularly successful. In particular, considering the so-called "diffusion limit" behavior of the evolving system (i.e. both spatially and temporally far from the source), the CTRW transport equation for a system undergoing anomalous diffusion in presence of an external advective field may be approximated by the asymptotic FADE model. This approximation is derived under the further assumption of truncating a required Laplace transform to the first non-constant term. The FADE model, by making use of the properties of fractional order derivatives, allows to obtain elegant close form analytical solutions. In this paper, by means of Monte Carlo simulation, we have explored the limits of validity of the FADE asymptotic solutions with respect to those of the CTRW, which must be in general obtained numerically. Both the case of a Galilei invariant and a Galilei variant advection schemes have been considered. Comparison with Monte Carlo results has revealed that the FADE solutions accurately represent the asymptotic behavior of the CTRW model only when $\alpha$ is far from its superior limit 1. As in most experimental evidences of anomalous kinetics $\alpha$ is usually close to 1 [see e.g. 8,18], we believe that the accuracy of the FADE solution should be improved. To this aim, we have expanded the cited Laplace transform by including higher order terms, whose importance has been shown to grow when $\alpha$ approaches 1. The proposed corrections do not allow in general to derive closed form solutions anymore, so that numerical inversion has been necessary: the solutions thus obtained turned out to be in very good agreement with the Monte Carlo solutions for CTRW, which have been adopted as reference curves. An appendix has been devoted to the derivation of the integral-differential equations corresponding to the improved Laplace transform expansions.

## Appendix A: Derivation of the Laplace transform of (2)

In order to obtain the Laplace transform of (2), we proceed as follows: by definition,

$$w(u) = \int_0^\infty e^{-ut} w(t)\, dt = \frac{2\alpha}{2+\alpha} \int_0^\tau e^{-ut} \frac{t}{\tau^2}\, dt + \frac{2\alpha}{2+\alpha} \tau^\alpha \int_\tau^{+\infty} e^{-ut} t^{-1-\alpha}\, dt\, .$$

By changing the variable of integration $ut = s$, we get

$$w(u) = \int_0^\infty e^{-ut} w(t)\, dt = \frac{2\alpha}{2+\alpha} \int_0^{u\tau} e^{-s} \frac{s}{u^2 \tau^2}\, ds + \frac{2\alpha}{2+\alpha} u^\alpha \tau^\alpha \int_{u\tau}^{+\infty} e^{-s} s^{-1-\alpha}\, ds\, .$$

Then, by integrating by parts, the first integral reads

$$\frac{2\alpha}{2+\alpha} \int_0^{u\tau} e^{-s} \frac{s}{u^2\tau^2} ds = \frac{2\alpha}{2+\alpha} \frac{1}{u^2\tau^2} \left(1 - e^{-u\tau}(1 + u\tau)\right).$$ Expanding the exponential in the diffusion limit $u\tau \ll 1$, we get:

$$\frac{2\alpha}{2+\alpha} \frac{1}{u^2\tau^2} \left(1 - e^{-u\tau}(1 + u\tau)\right) \approx \frac{\alpha}{2+\alpha} - \frac{2}{3} \frac{\alpha}{2+\alpha} (u\tau).$$

Then, integrating twice by parts, the second integral reads:

$$\frac{2\alpha}{2+\alpha} u^\alpha \tau^\alpha \int_{u\tau}^{+\infty} e^{-s} s^{-1-\alpha} ds = \frac{2}{2+\alpha} \left( e^{-u\tau} + \frac{(u\tau)e^{-u\tau}}{1-\alpha} - \frac{1}{1-\alpha} \int_{u\tau}^{+\infty} e^{-s} s^{1-\alpha} ds \right).$$

In the diffusion limit $u\tau \ll 1$, expanding the exponential and recalling that $\lim_{u\tau \to 0} \int_{u\tau}^{+\infty} e^{-s} s^{1-\alpha} ds \sim \Gamma(2-\alpha)$, we thus obtain 
$$\frac{2}{2+\alpha} \left( e^{-u\tau} + \frac{(u\tau)e^{-u\tau}}{1-\alpha} - \frac{1}{1-\alpha} \int_{u\tau}^{+\infty} e^{-s} s^{1-\alpha} ds \right) \approx \frac{2}{2+\alpha} \left(1 + \frac{\alpha}{1-\alpha}(u\tau)\right) - \Gamma(1-\alpha)(u\tau)^\alpha.$$

Suitable simplifications finally lead to $w(u) = 1 - \frac{2\Gamma(1-\alpha)}{2+\alpha}(u\tau)^\alpha + \left(\frac{2\alpha}{(2+\alpha)(1-\alpha)} - \frac{2}{3}\frac{\alpha}{2+\alpha}\right)(u\tau).$

**Appendix B: Derivation of equations (19a) and (20a)**

Let us consider the Chapman-Kolmogorov relation (1) and rewrite this equation as follows:

$$u P(k,u) = 1 - w(u) + w(u) \lambda(k) P(k,u) \tag{B1}$$

After some algebra, the terms may be arranged in the following way:

$$u P(k,u) - 1 = \frac{u\, w(u)}{1 - w(u)} \left(- P(k,u) + \lambda(k) P(k,u)\right) \tag{B2}$$

By putting $\Psi(u) = \frac{u\, w(u)}{1 - w(u)}$ and inverse Laplace transforming, we get (recalling that $P(k,0)=1$):

$$\frac{\partial}{\partial t} P(k,t) = \int_0^t \Psi(t-t') \left(- P(k,t') + \lambda(k) P(k,t')\right) dt'. \tag{B3}$$

Then, inverse Fourier transforming, we get

$$\frac{\partial}{\partial t} P(x,t) = \int_0^t \Psi(t-t') \left(- P(x,t') + \int_{-\infty}^{+\infty} \lambda(x - x') P(x',t') dx'\right) dt' \tag{B4}$$

By applying to $\lambda(.)$ the properties of the expansion of a distribution in delta function and its derivatives up to the second order, provided that the first and second moments exist and are finite [36], we get:

$$\lambda(x - x') = \delta(x - x') + \langle \lambda(x') \rangle \delta'(x - x') + \frac{1}{2} \langle \lambda^2(x') \rangle \delta''(x - x'). \tag{B5}$$

Then, we make use of the definition of the *n-th* order derivative of a delta function (for any sufficiently well-behaved $g(x')$), viz.

$$\int_{-\infty}^{+\infty} \delta^n(x-x')\,g(x')\,dx' = (-1)^n \int_{-\infty}^{+\infty} \delta(x-x')\frac{d^n}{dx'^n}g(x')\,dx' = (-1)^n \frac{d^n}{dx'^n}g(x')\bigg|_{x'=x}. \quad (B6)$$

Finally, recalling the normalization property of $\lambda(x')$ and suitably simplifying, we obtain:

$$\frac{\partial}{\partial t}P(x,t) = \int_0^t \Psi(t-t')\left(\frac{1}{2}\langle \lambda^2(x')\rangle \frac{\partial^2}{\partial x^2}P(x,t') - \langle \lambda(x')\rangle \frac{\partial}{\partial x}P(x,t')\right)dt'. \quad (B7)$$

Expression (B7) has the structure of a Fokker-Planck-Kolmogorov equation (FPK) and can be directly derived – as shown – from the Chapman-Kolmogorov equation (B1) under the only assumption of arresting expansion (B5) to the second order. Formulation (B7) is very general and contains as particular cases FDE (5), FADE (8) and (15) and integral-differential equations (19a) and (20a), depending on the assumptions which are made on its structure. In particular, if we assume that $\langle \lambda(x')\rangle = 0$ and we truncate expansion (3) to the first non-constant term, so that $\Psi(u) = \frac{u^{1-\alpha}}{c_\alpha \tau^\alpha}$ we recover the FDE (5). If we then add to the previous hypotheses the Galilei invariance property, we recover the FADE (8). When on the contrary we substitute (3) with (18), the same assumptions lead to equation (19a). On the other hand, if we assume that the jump lengths pdf has a finite mean $\langle \lambda(x')\rangle = \mu$ and expansion (3) holds true we are led to the Galilei variant FADE (15), while substituting under the same hypothesis (3) with (18) allows to obtain equation (20a). It has been always assumed that $\langle \lambda^2(x')\rangle = 2\sigma^2$.

The integral kernel $\Psi(t)$ can be regarded as a memory kernel which accounts for the past history of the process $P(x,t)$, from the initial time $t=0$ to the present, while the operator $\frac{\partial}{\partial t}$ represents the dependence from local time [18,36]. As such, equation (B7) is *a priori* non-Markovian. Moreover, $\Psi(t)$ depends only on the particular functional form of the waiting times distribution $w(t)$: in particular, if we assume $w(u) = \frac{1}{1+u\tau}$, i.e. $w(t) = \frac{1}{\tau}e^{-\frac{t}{\tau}}$, $\Psi(u) = \tau^{-1}$, so that $\Psi(t-t') = \frac{1}{\tau}\delta(t-t')$ and the general form of the Fokker-Planck-Kolmogorov equation (B7) reduces to:

$$\tau \frac{\partial}{\partial t}P(x,t) = \frac{1}{2}\langle \lambda^2(x')\rangle \frac{\partial^2}{\partial x^2}P(x,t) - \langle \lambda(x')\rangle \frac{\partial}{\partial x}P(x,t) \quad (B8)$$

which is but the well-known Markovian Advection-Dispersion Equation (ADE), without memory[1]. This is intimately connected to the exponential pdf assumed for the waiting times, as for this distribution the transition rate, i.e. the probability per unit time to effectuate a jump in a *dt* after time *t* – given that up to *t* no jump has been elapsed –, is a constant $\forall t$ and does not depend on the past history [18,21]. We remark that, in this case, the Galilei variant and the Galilei invariant schemes formally become indistinguishable. Moreover, in equation (B8) $\frac{1}{2}\frac{\langle \lambda^2(x')\rangle}{\tau} = D^*$ may be interpreted as the dispersion coefficient, while $\frac{\langle \lambda(x')\rangle}{\tau} = v^*$ represents the advection coefficient.

We give now a proof of the expansion (B5) and present some remarks on its physical meaning. We begin by introducing a sufficiently well-behaved test function $f(x)$, e.g. $f(x) \in C^\infty(\Re)$, and we integrate the product of $f(x)$ and of the r.h.s. of (B5) over $dx'$. By making use of the properties of the delta function derivatives (B6), we obtain:

$$\int \delta(x-x')f(x')\,dx' + \langle \lambda(x')\rangle \int \delta'(x-x')f(x')\,dx' + \frac{1}{2}\langle \lambda^2(x')\rangle \int \delta''(x-x')f(x')\,dx' =$$

---
[1] It is interesting to remark that $\Psi(u)$ is dimensionally the inverse of a time. The constant $\tau$ may be regarded as a characteristic time scale of the stochastic process.

$$= f(x) - \langle \lambda(x') \rangle f'(x) + \frac{1}{2} \langle \lambda^2(x') \rangle f''(x), \tag{B9}$$

where the apex of *f* denotes derivation with respect to its argument. Then, we integrate the product of l.h.s. of (B5) and of *f(x)* over *dx'*, obtaining

$$\int \lambda(x-x') f(x') dx'. \tag{B10}$$

Let now expand the function *f(x')* in Taylor series around the point *x*, viz. $f(x') = \sum_{m=0}^{+\infty} \frac{f^{(m)}(x)}{m!} (x'-x)^m$. If this expansion is arrested to the second order, i.e. to *m=2*, expression (B10) becomes

$$f(x) \int \lambda(x-x') dx' + f'(x) \int \lambda(x-x')(x'-x) dx' + \frac{f''(x)}{2} \int \lambda(x-x')(x'-x)^2 dx' =$$

$$= f(x) - f'(x) \langle \lambda(x') \rangle + f''(x) \frac{\langle \lambda^2(x') \rangle}{2}, \text{ by taking into account the normalization of } \lambda(x').$$

This expression for the l.h.s. (B10) coincides with the one for the r.h.s. (B9) and thus proves identity (B5).

The physical meaning of a second order expansion in expression (B5) may be understood in the following terms. Let the test function *f(x')* be $f(x') = e^{ikx'} \in C^\infty(\Re)$ and put *x=0*, without loss of generality. Then, the l.h.s. of identity (B5) becomes

$$\int_{-\infty}^{+\infty} \lambda(-x') e^{ikx'} dx' = \int_{-\infty}^{+\infty} \lambda(x') e^{-ikx'} dx' = \mathcal{F}\{\lambda(x'), k\} \tag{B11}$$

which is by definition the Fourier transform (in the reciprocal variable *k*) of the function *λ(x')*.
Analogously, by making use of the properties of the delta function and of its derivatives, the r.h.s. of (B5) becomes

$$\int \delta(-x') e^{ikx'} dx' + \langle \lambda(x') \rangle \int \delta'(-x') e^{ikx'} dx' + \frac{1}{2} \langle \lambda^2(x') \rangle \int \delta''(-x') e^{ikx'} dx' =$$

$$= 1 + \langle \lambda(x') \rangle ik - \frac{1}{2} \langle \lambda^2(x') \rangle k^2. \tag{B12}$$

Thus, the assumption of a second order expansion in expression (B5) coincides with the following approximation in the Fourier reciprocal space:

$$\mathcal{F}\{\lambda(x'), k\} = 1 + \langle \lambda(x') \rangle ik - \frac{1}{2} \langle \lambda^2(x') \rangle k^2, \tag{B13}$$

which is but the usual diffusion limit expansion, which holds true provided that $k\sqrt{\frac{1}{2}\langle \lambda^2(x') \rangle} \ll 1$ or analogously $x' \gg \sqrt{\frac{1}{2}\langle \lambda^2(x') \rangle}$ in the direct space.